\documentclass[fleqn,usenatbib]{mnras}
\usepackage{newtxtext,newtxmath}
\usepackage[T1]{fontenc}
\usepackage{ae,aecompl}
\usepackage{tabularx}
\usepackage{graphicx}
\usepackage{float}
\usepackage{mathrsfs}	
\usepackage{amsmath}
\usepackage{color}
\usepackage{hyperref}
\usepackage{multirow}
\usepackage{subfigure}
\usepackage{longtable}

\usepackage{graphicx}	% Including figure files
\usepackage{amsmath}	% Advanced maths commands
\usepackage{amssymb}	% Extra maths symbols
\usepackage{csvsimple}

\title [search for galaxy-Ly$\alpha$ emitter lens candidates]{Using deep Residual Networks to search for galaxy-Ly$\alpha$ emitter lens candidates based on spectroscopic-selection} 
\author[Li et al.]
{Rui Li$^{1,2,3,4}$\thanks{E-mail: lirui@ynao.ac.cn}, Yiping Shu$^{5,6}$\thanks{E-mail: yiping.shu@pmo.ac.cn}, Jianlin Su$^{7}$, Haicheng Feng$^{1,2,3,4}$,Guobao Zhang$^{1,2,3,4}$, 
\newauthor Jiancheng Wang$^{1,2,3,4}$\thanks{E-mail: jcwang@ynao.ac.cn}, Hongtao Liu$^{1,2,3,4}$
\vspace*{0.2cm}\\
$^{1}$ Yunnan Observatories, Chinese Academy of Sciences, 396 Yangfangwang, Guandu District, Kunming, 650216, P. R. China\\ 
$^{2}$ University of Chinese Academy of Sciences, Beijing, 100049, P. R. China \\
$^{3}$Center for Astronomical Mega-Science, Chinese Academy of Sciences, 20A Datun Road, Chaoyang District, Beijing, 100012,\\ 
\ \ \ P. R. China \\ 
$^{4}$ Key Laboratory for the Structure and Evolution of Celestial Objects, Chinese Academy of Sciences, 396 Yangfangwang, Guandu \\
\ \ \ District, Kunming, 650216, P. R. China\\
$^{5}$ Purple Mountain Observatory, Chinese Academy of Sciences, 2 West Beijing Road, Nanjing, Jiangsu, 210008, China \\
$^{6}$ Institute of Astronomy, University of Cambridge, Madingley Road, Cambridge CB3 0HA, UK\\
$^{7}$ School of Mathematics, Sun Yat-sen University, Guangzhou, China}

\date{Accepted XXX. Received YYY; in original form ZZZ}

\pubyear{2018}

\begin{document}
\label{firstpage}
\pagerange{\pageref{firstpage}--\pageref{lastpage}}
\maketitle

% Abstract of the paper
\begin{abstract}
More than one hundred galaxy-scale strong gravitational lens systems have been found by searching for 
the emission lines coming from galaxies with redshifts higher than the lens galaxies. Based
on this spectroscopic-selection method, we introduce the deep Residual Networks (ResNet, a kind
of deep Convolutional Neural Networks) to search for the galaxy-Ly$\alpha$ emitter (LAE) lens candidates by recognizing the Ly$\alpha$ emission lines coming from high redshift galaxies ($2 < z < 3$) in the spectra of early-type galaxies (ETGs) at middle redshift ($z\sim 0.5$). The spectra of the ETGs come from the Data Release 12 (DR12) of the Baryon Oscillation Spectroscopic Survey (BOSS) of the Sloan Digital Sky Survey \uppercase\expandafter{\romannumeral3} (SDSS-\uppercase\expandafter{\romannumeral3}). In this paper, we first build a 28 layers ResNet model, and then artificially synthesize 150,000 training spectra, including 140,000 spectra without Ly$\alpha$ lines and 10,000 ones with Ly$\alpha$ lines, to train the networks. After 20 training epochs, we obtain a near-perfect test accuracy at 0.9954. The corresponding loss is 0.0028 and the completeness is 93.6\%. We finally apply our ResNet model to our predictive data with 174 known lens candidates. We obtain 1232 hits including 161 of the 174 known candidates 	(92.5\% discovery rate). Apart from the hits found in other works, our ResNet model also find 536 new hits. We then perform several subsequent selections on these 536 hits and present 5 most believable lens candidates.
\end{abstract}

% Don't make up new ones.

%%%%%%%%%%%%%%%%% BODY OF PAPER %%%%%%%%%%%%%%%%%%
\section{Introduction}

Galaxy-scale strong gravitational lens, which can provide very tight constraints on the projected mass of the lens galaxies, is not only an available probe to study the structures, formation and evolution of galaxies \citep[e.g.][]{2006ApJ...649..599K, 2010ApJ...724..511A, 2013ApJ...777...98S, 2012ApJ...757...82B}, but also a crucial technique to study cosmology \citep{2013ApJ...766...70S}. %made an blind analysis to the time-delay lens system RX J1131-1231 and successfully quantified the Hubble constant $H_0$ with lower uncertainty ($H_{0}=78.7^{+4.3}_{-4.5}$).
 Additionally, gravitational lenses could act as natural telescopes and give us magnified views of the background objects, which could help us to study the celestial bodies in the distant universe   \citep[e.g.][]{2008Natur.456..927I, 2009MNRAS.400.1121S, 2015ApJ...812..114T}.
 
Several methods have been used to search for the galaxy-scale strong gravitational lenses. Among them, an efficient one is searching for the arc-like morphological features of strong lens systems \citep{2004A&A...416..391L, 2005ApJ...633..768H, 2007A&A...472..341S, 2008MNRAS.385..918K, 2012ApJ...749...38M}. In order to find strong lens systems in a great deal of galaxy images. , \cite{2014ApJ...785..144G} subtracted the light of the central galaxies using multi-band images and then searched for the arc-like shapes in the residual images. Another modified work was done by \cite{2014A&A...566A..63J}, who successfully subtracted the light of the foreground galaxies with Principal Component Analysis (PCA), and then searched for gravitational lens features in the residual images. They pointed out that the PCA-based galaxy subtraction algorithm performed better than the traditional model fitting method\citep{2014ApJ...785..144G}.

Besides the method of morphological recognition, \cite{1996MNRAS.278..139W} proposed a spectroscopic-selection technique to search for galaxy-galaxy strong lens systems. This technique identifies lens candidates by looking for compound spectra. As shown in Figure \ref{fig:real_arti}, the left panel is a compound spectrum of an identified lensing system \citep[SDSSJ 1110+2808,][]{2016ApJ...833..264S}. The foreground galaxy of this lens system is an early-type galaxy (ETG,  $z\sim 0.6073$).  Between 3800\r{A} and 5000\r{A}, there is no emission line in the spectrum. However, near 4150\r{A}, we find a Ly$\alpha$ line which comes from a Ly$\alpha$ emitter (LAE). The LAE is at the redshift of $z=2.3999$ and lies close to the line of sight of the foreground ETG. The light from the LAE is deflected by the foreground ETG, and then several images of the background LAE emerge around the foreground galaxy. Further detailed description of the spectroscopic-selection technique can be found in \cite{2004AJ....127.1860B}. In general, this technique identifies lens candidates by searching for the spectra with emission lines from background sources.

Several galaxy-scale strong lens surveys have been made with the above spectroscopic-selection technique, such as the Sloan Lens ACS  Survey \citep[SLACS,][]{2006ApJ...638..703B, 2008ApJ...682..964B}, the Sloan WFC Edge-on Late-type Lens Survey \citep[SWELLS,] []{2011MNRAS.417.1601T}, the  Baryon Oscillation Spectroscopic Survey (BOSS) Emission-Line Lens Survey \citep[BELLS,][]{2012ApJ...757...82B}, the SLACS for the Masses \citep[S4TM,][]{2015ApJ...803...71S} and the BELLS for the GALaxy-Ly$\alpha$ EmitteR sYstems Survey \citep[BELLS GALLERY,][]{2016ApJ...833..264S}. More than one hundred galaxy-galaxy strong gravitational lenses have been found in these surveys.  A lot of  significant works about the structures, formation and evolution of ETGs have been done with these lens systems \citep{2006ApJ...649..599K, 2009ApJ...705.1099A, 2012ApJ...757...82B, 2015ApJ...803...71S}. Recently, using a similar technique, \cite{2017arXiv171101184M} have found 9 secure QSO-Galaxy lens candidates, presenting the potential application of the spectroscopic-selection technique in QSO Lens searching. So far, this spectroscopic-selection technique is one of the most efficient methods. However, the speed of this traditional method is not fast enough, especially in the coming age of the big data in astronomy. 

Resently, the Convolutional Neural Networks (CNNs), a kind of machine learning algorithm, have been applied in astronomy. The first application of the CNNs is the spectral classification of the data release 10 (DR10) of the Sloan Digital Sky Survey(SDSS). \cite{2014arXiv1412.8341H} applied his CNN model on more than 60000 spectra of SDSS and yielded a success rate of nearly 95\%. Their work conclusively proved the great potential of the CNNs in astrophysics. Later, the CNNs have been applied in some other works, such as the morphological classification of SDSS galaxies \citep{2015MNRAS.450.1441D}, the photometric redshift estimates of SDSS galaxies \citep{2016A&C....16...34H} and the star/galaxy classification \citep{2017MNRAS.464.4463K}. In the field of gravitational lenses, \cite{2017MNRAS.472.1129P} and \cite{2018MNRAS.473.3895L} have used the CNNs to recognize galaxy-scale strong gravitational lenses based on morphological classification. In \cite{2017MNRAS.472.1129P}, they found 761 strong lens candidates in the Kilo Degree Survey (KiDS) and presented 56 most reliable candidates after manual selection.

The CNNs have not been applied for searching gravitational lens candidates based on the spectroscopic-selection technique. In this paper, we will use the deep Residual Networks (ResNet, one popular model of CNNs) to search for the galaxy-Ly$\alpha$ emitter lens candidates in DR12 of SDSS by recognizing the high redshift ( $2 < z < 3$) Ly$\alpha$ emission lines.  LAEs are a kind of young, low-mass galaxies with an extremely high star-formation rate. They are important components of the early universe, and critical for studying the formation and evolution of galaxies. Benefited from the magnifying effect, galaxy-LAE lenses can provide us good opportunities to study the LAEs themselves. More importantly, galaxy-LAE lenses can help us to exploit the small-scale dark substructures around the lens galaxies and constrain the slope and normalization of the substructure-mass function \citep{2016ApJ...824...86S}. 

The paper is organized as follows. In Section 2, we provide a brief description of Machine Learning, CNNs and ResNet. In Section 3, we talk about how to use ResNet to recognize emission lines that are not from the foreground galaxies themselves. In Section 4, we present 5 new gravitational lens candidates found by our ResNet model. In Section 5, we discuss the advantages and the possible improvement of our strategy.

\section{Machine Learning, CNNs and ResNet}
In this section, we provide a brief introduction of Machine Learning, CNNs and ResNet.

\begin{figure*}
\centering
\includegraphics[height=0.28\textwidth, width=0.90\textwidth]{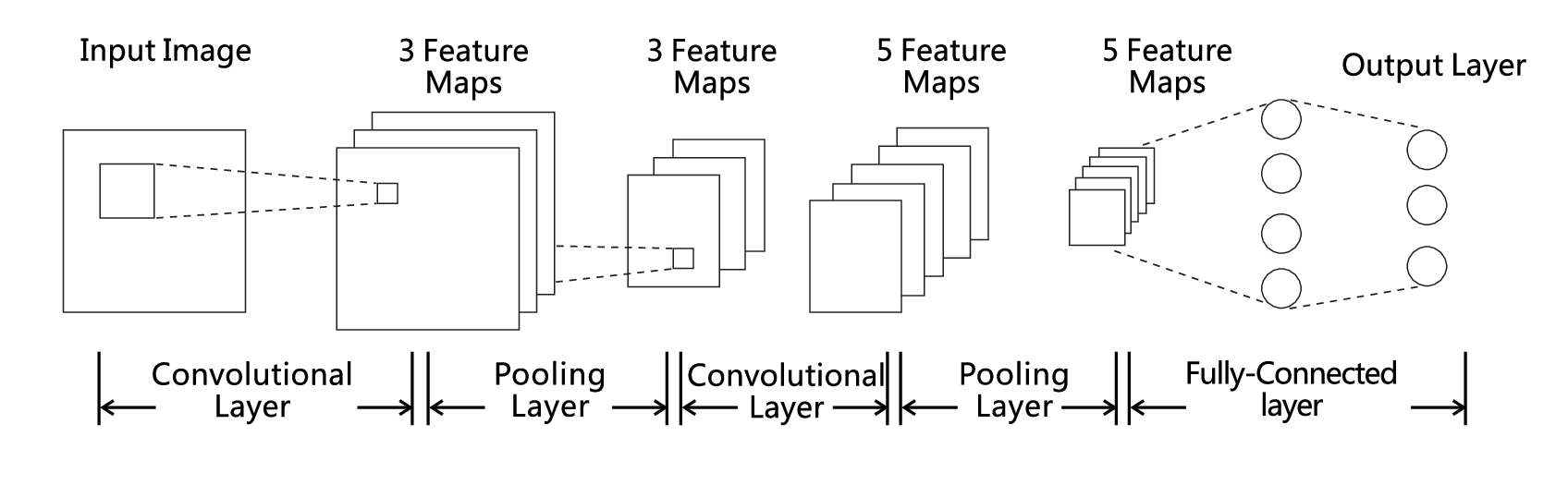}
\caption{\label{fig:connect}
A simple but complete CNN model with two convolutional layers, two pooling layers and one Fully-connected layers.}
\end{figure*}

\subsection{The Machine Learning and Deep Learning}
Machine Learning was coined in 1959 by \cite{1959IBM...3.3}. It is a field of computer science that gives computers the ability to learn from train data and then make predictions on predictive data. Depending on whether there are given ``labels" for  the learning systems, Machine Learning is typically classified into two classes, supervised learning and  unsupervised learning. For supervised learning, the train data as well as the ``labels'' for the train data are given to the learning algorithm.  Then the learning algorithm would ``learn'' a general rule that maps inputs to outputs. For unsupervised learning, no ``labels'' are given, the learning algorithm has to discovery hidden patterns in the train data by itslef.  The Machine Learning  algorithm could be designed to achieve various tasks, such as classification, regression,  clustering, density estimation and so on. The task in our work is in fact a classification task which can be tackled in a supervised way. The train data in our work could be divided into two classes and given different ``labels''. Then the learning algorithm could ``learn'' from the train data and their ``labels'' then produce a model that assigns the inputs to one of the two classes.

Deep learning is a kind of Machine Learning with multiple layers to extract the feature of the inputs. Each later layer uses the outputs of the previous layer as inputs. Most deep learning models are based on artificial neural networks (ANNs) which consist of many artificial neurons. Similar with biophysical neurons, the artificial neurons of ANNs can receive inputs, change their states according to the inputs, and then produce outputs depending on their states. If the artificial neurons of deep ANNs can only receive informations from the previous layer and output informations to the next layer, we call this kind of ANNs as deep feedforward neural networks. There are no cycles or loops in this kind networks. More detailed information about Deep learning can be found in \cite{2018MNRAS.473.3895L}.

\subsection{The CNNs}
\begin{figure}
\centering
\includegraphics[height=0.23\textwidth, width=0.36\textwidth]{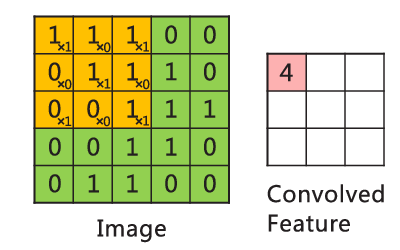}
\caption{\label{fig:convol}
Illustration of the convolutional layer. A $5\times 5$ image (green) is convolved by a $3\times 3$ convolutional kernel (yellow). The result of this convolutional operation is a $3\times 3$ feature map. The convolutional kernel moves on the image by 1 stride and then output the convolutional result to the feature map.}
\end{figure}

\begin{figure}
\centering
\includegraphics[height=0.22\textwidth, width=0.44\textwidth]{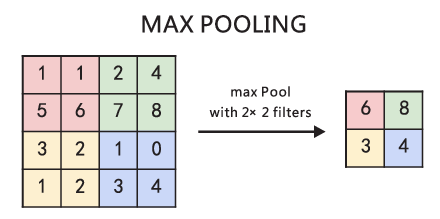}
\caption{\label{fig:pool}
Illustration of the pooling layer. It is a max pooling layer with $2\times 2$ filters. The filters will find the maximum values of the corresponding local areas and then use the maximum values to represent these local areas to form a new compressed feature map.}
\end{figure}

As one of the most popular models of deep learning, recent years, the CNNs have gained a lot of attentions in image recognition, speech recognition, motion analysis, nature language processing, as well as many other researching fields. CNNs are a kind of deep feedforward neural networks and formed by a stack of distinct layers. The layers mainly contain the convolutional layers, the pooling layers and the fully-connected layers. Here are some important terminologies of CNNs used in this paper:

\begin{enumerate}
\item {\bf Convolutional layers:} The convolutional layers apply convolutional operations to the inputs and then pass the results to the next layers. Figure \ref{fig:convol} (from the internet, just for illustration) shows a simple convolutional operation. A $5\times 5$ input image (green) is convolved by a $3\times 3$ convolutional kernel (yellow,  also known as convolutional neurons). The result of this convolutional operation is a $3\times 3$ feature map. Simply speaking,  the convolutional layers are used to extract different features of the inputs. Usually, there are several convolutional neurons in one feature map. The convolutional neurons are composed of weights and bias which could be written as ${\bf \lbrace W, b\rbrace}$. The weights and bias could be trained by the train data.
\item {\bf Pooling layers}: The pooling layers are introduced between two convolutional layers (local pooling layers) or before the Fully-connected layers (global pooling layers) to compress the feature maps and simplify the calculations. There are two common pooling methods, max pooling and average pooling. Figure \ref{fig:pool} (from the internet, just for illustration) shows a max pooling with $2\times 2$ filters. Each filter could find the maximum value of the corresponding local area to represent this local area in order to form a new compressed feature map.
\item {\bf  Fully-connected layers:}  The fully-connected layers connect every neuron in one layer to every neuron, in another layer to turn all the local feature maps into a global feature map.
\item {\bf Dropout layers:} All of the parameters of the weights and bias would occur in the fully-connected layers, which probably cause overfitting. One method to reduce the overfitting is to introduce the Dropout layers. During each training process, we can randomly ``drop out'' some parameters of ${\bf \lbrace W, b\rbrace}$ with probability $1-p$ or keep the parameters with probability $p$. After a ``Dropout'' operation, a reduced network is left and would be tained on data.
\item{\bf Activation function:} The Activation function defines the outputs of a neuron given inputs. A neuron receives signals and ``activate'' them using Activation function, then output the activated signals to the next neuron. For CNNs, only non-linear Activation functions could be used, because if the Activation functions are linear, there were only linear transformations between two layers, which could reduce the capacity of the networks. Additionally, non-linear Activation functions can limit the outputs to be finite values \citep{2016arXiv160300391G}.
\item {\bf Loss:} A Machine Learning model is trained by minimising a given loss that measure the deviation between the predicted values and the real values. The loss function could be defined by the user.
\item{\bf Accuracy and Completeness:} The accuracy and completeness are used to measure the capacity of a classification model.  For example, we have totally $m$ samples, among them, $n$ samples are positive samples, the remain are negative. If $i$ samples are classified as positive by our model, among them, $j$ samples are real positive. Then the accuracy is defined as $j/i$, the completeness is defined as $j/n$.
\item {\bf Epoch:} Epoch is a terminology of Keras\footnote{https$://$keras.io} used in this work. Keras is a high-level neural networks Application Programming Interface (API), and capable of running on top of TensorFlow\footnote{https://github.com/tensorflow/tensorflow}, CNTK\footnote{https://www.microsoft.com/en-us/cognitive-toolkit/}, or Theano\footnote{http://deeplearning.net/software/theano/}. For Keras, the train dataset would be divided into lots of batches, the networks are updated when a batch is processed. An epoch is an iteration over the entire dataset.
\end{enumerate}

The entire training process of a CNN model is as follow: $1\rangle$ Initialize the weights and bias ${\bf \lbrace W, b\rbrace}$ of the CNN model. $2\rangle$ The inputs are convolved by the convolutional layers, resulting in some feature maps. Then the feature maps are compressed by the pooling layers. Repeat this process if necessary.  $3\rangle$ At the end of the netwoks, the fully-connected layers are used to obtain the predicted values. $4\rangle$ Calculate the deviation between the predicted values and the input values. If the deviation is bigger than our expectation, update the weights and bias ${\bf \lbrace W, b\rbrace}$ according to the deviation and retrain the networks. Figure \ref{fig:connect} (from the internet, just for illustration) shows a simple but complete CNN model with two convolutional layers, two pooling layers and one fully-connected layer. The input image is convolved by 3 convolutional kernels, resulting in 3 feature maps. Then these 3 feature maps are compressed by a pooling layer. After another convolutional layer and pooling layer, a fully-connected layer is used to turn the local feature maps into a global feature map and output the prectived values.

\subsection{The ResNet}
Many evidences show that the network depth is very important for the classification and regression tasks. Theoretically, the training results of the networks should become better as the layers become deeper. However, for deep networks, a ``degradation" problem would occur when they start converging. With the increase of network depth, the accuracy gets saturated and then degrades rapidly \citep{2015arXiv151203385H}, which is not caused by over fitting. That's why many popular CNN models have only a few layers \citep[e.g., 5 layers in LeNet, 7 layers in AlexNet;][]{1998IEEE,2012ANIPS}. In order to overcome the degradation of the accuracy of deep networks, \cite{2015arXiv151203385H} proposed a new CNN model: the deep Residual Networks.

\begin{figure}
\centering
\includegraphics[height=0.25\textwidth, width=0.37\textwidth]{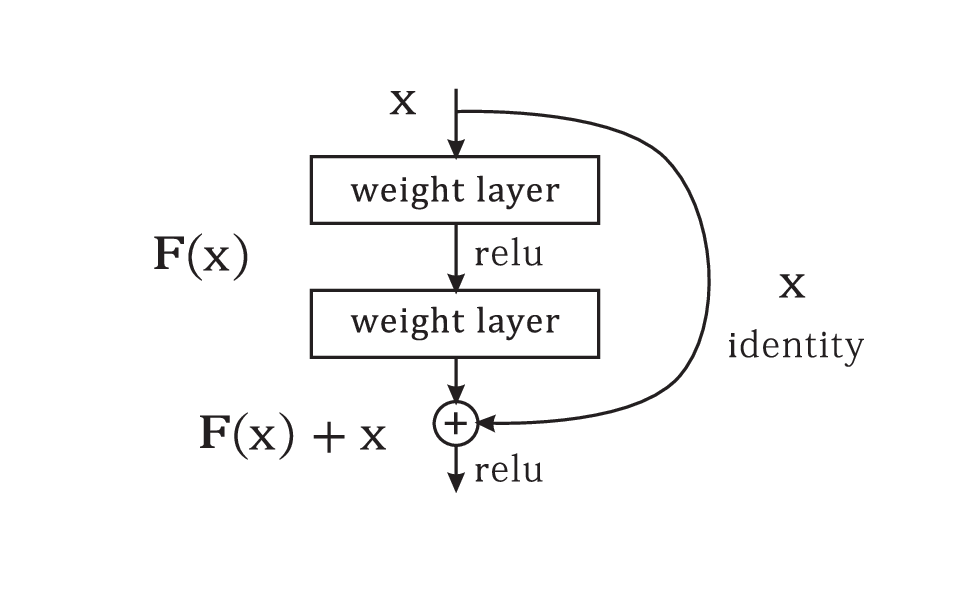}
\caption{\label{fig:ResNet_block}
Residual learning: a building block with a shortcut connection \citep{2015arXiv151203385H}. The weight layers are the assemble of several convolutional layers and other necessary layers. Relu \citep{2011AISTATS} is one kind of Activation Function. }
\end{figure}

ResNet assumes that it is easier to optimize the residual mapping than to optimize the origin, unreferenced mapping. Denoting the desired underlying mapping as $H(x)$ which would be fitted by a few stacked layers. $x$ is the inputs of the first layer. ResNet lets the stacked layers to fit the residual mapping of $F(x):=H(x)-x$. Then the original mapping $H(x)$ becomes $F(x)+x$, and it could be realized by the networks with ``shortcut connections" (see Figure \ref{fig:ResNet_block}). The ``shortcut connections" simply perform the identity mapping, then skip one or more layers and add their outputs to the outputs of the stacked layers. In \cite{2015arXiv151203385H}, the authors explained that ``With the residual learning reformulation, if identity mappings are optimal, the solvers may simply drive the weights of the multiple nonlinear layers toward zero to approach identity mappings."

In our study, we use Machine Learning to find gravitational lens candidates based on spectroscopic-selection. Spectrum is one dimensional sequence that has local correlations. We  use one dimensional CNNs to build the Machine Learning model because CNNs have good performance in distinguishing the inputs with local correlations. In order to prevent the ``degradation" problem of deep networks and reduce the training difficulty, we adopt the ResNet model to our study. In fact, before the final choice of ResNet, we have tried several ``plain" networks (e.g., LeNet; AlexNet), but they are difficult to converge in our study.  

\section{Find emission lines with ResNet}

\begin{figure*}
\includegraphics[height=4.9cm,width=8.5cm]{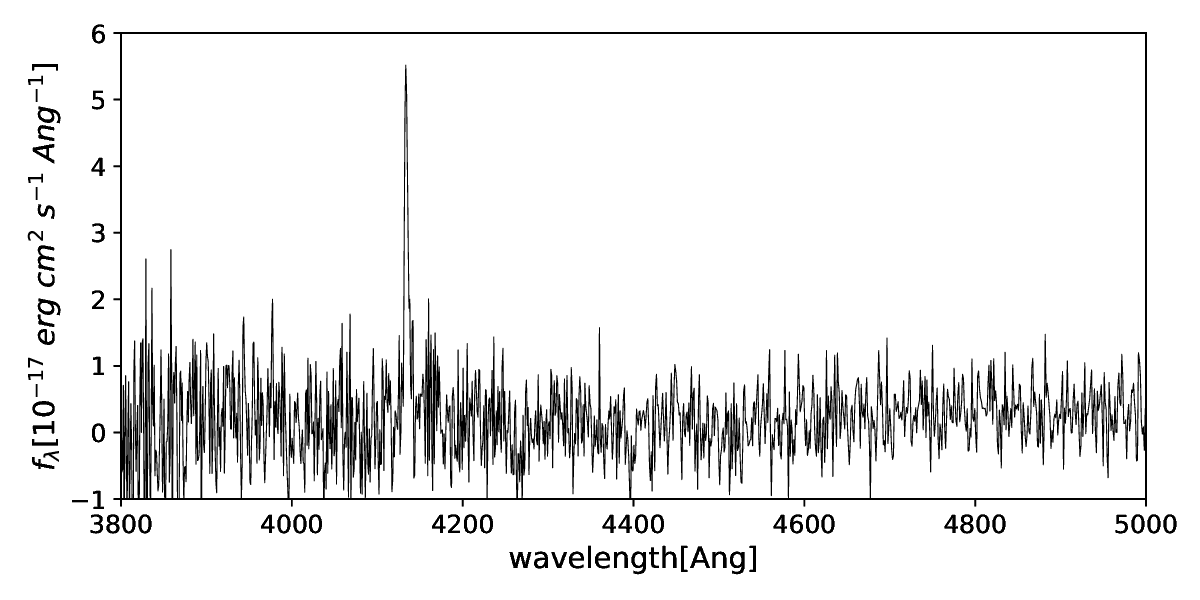}
\includegraphics[height=4.9cm,width=8.5cm]{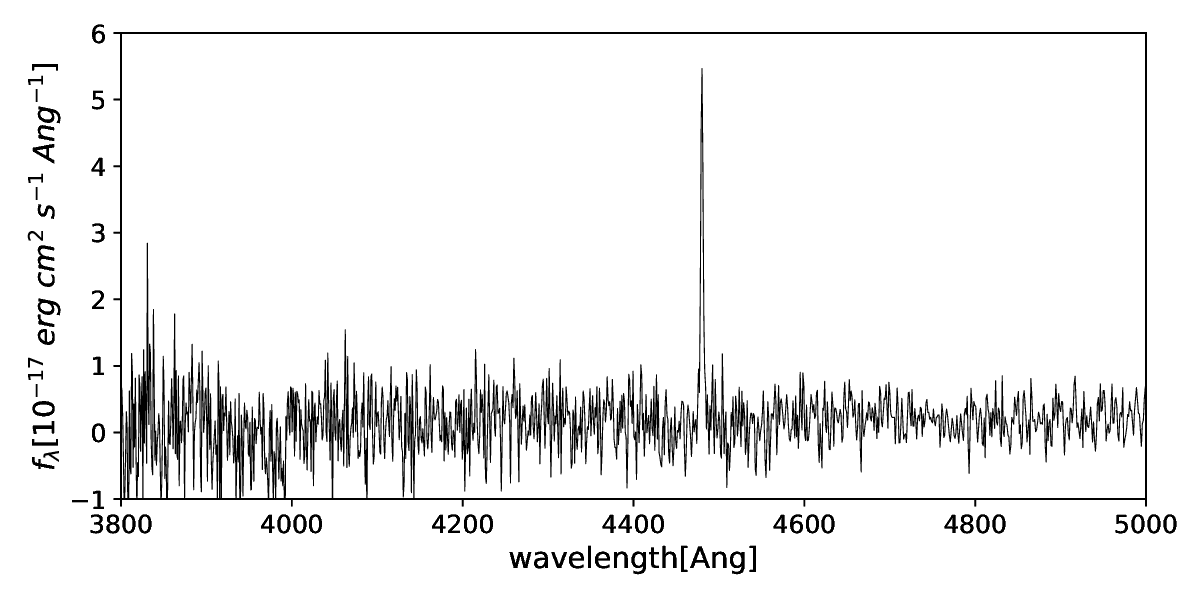}
\caption{\label{fig:real_arti}
The left is the spectrum of SDSSJ 1110+2808 from BELLS GALLERY. The right is an artificial one. Both have high SNR Ly$\alpha$ lines.}
\end{figure*}

So far, there are several gravitational lens surveys based on spectroscopic-selection (e.g., the SLACS, BELLS, S4TM and BELLS GALLERY). The SLACS, BELLS and S4TM concentrated on searching for multiple emission lines (meanly the [O{\uppercase\expandafter{\romannumeral2}}] 3727 and [O{\uppercase\expandafter{\romannumeral3}}] 5007) from the galaxies with redshift higher than the foreground galaxies, while the BELLS GALLERY focused on only one Ly$\alpha$ line. We put forward that Machine Learning could be used to search for galaxy scale gravitational lens candidates based on spectroscopic-selection. As the first apllication, in this study we use ResNet to find some galaxy-LAE lens candidates in the BOSS spectra by searching for high redshift ($2 < z < 3$)  Ly$\alpha$ lines in the spectra of middle redshift ($ z \sim 0.5$) galaxies. However, in the band used in this work (3800\r{A} $<\lambda<$ 5000\r{A}, see Section 3.2 ), the spectra with lower redshift [O{\uppercase\expandafter{\romannumeral2}}] 3727 or other emission lines could also be found by our networks. We will abandon these spectra using several other subsequent selections (see Section 4).

\subsection{The BELLS GALLERY survey}
The lens candidates of BELLS GALLERY are selected from the DR12 of the BOSS by searching for Ly$\alpha$ emission lines from high redshift Ly$\alpha$ emitters, and then following up high-resolution Hubble Space Telescope (HST) imaging observations to confirm the lens nature. In \cite{2016ApJ...824...86S}, the first and most important step is selecting out the spectra with emission lines that are not from the target galaxies themselves. To do this, they confined their research to the observed wavelength range of 3600\r{A} $<\lambda<$ 4800\r{A} (roughly $2 < z < 3$ for the Ly$\alpha$ lines) and used an error-weighted matched-filter with a Gaussian kernel to search for the Ly$\alpha$ lines in BOSS targets. This main step yielded 4982 hits. Then they performed several subsequent selections to reject the false hits and finally found 187 lens candidates with obvious Ly$\alpha$ emission lines. They presented 21 high quality BELLS GALLERY candidates in their paper. Among them 17 candidates have been confirmed as strong gravitational lenses by HST imaging observations. The goal of the BELLS GALLERY survey is exploiting the small-scale dark substructure around ETGs and constraining the slope and normalization of the substructure-mass function \citep{2016ApJ...824...86S}. 

BELLS GALLERY is a successful lens candidates searching work. Whereas, the first step of the search is complex and consume too much time. First, they have to obtain the accurate spectral flux errors across the observed spectra. Second, they need to fit each spectrum with standard ETGs spectroscopic templates to obtain the galaxy-subtracted residual spectrum. For millions of targets, these procession would consume a lot of time. In order to increase the searching speed, we introduce ResNet to automatically select out the spectra with emission lines that are not from the targets themselves. 

\subsection{The predictive data and the train/test data}

\begin{figure*}
\includegraphics[height=5cm,width=8cm]{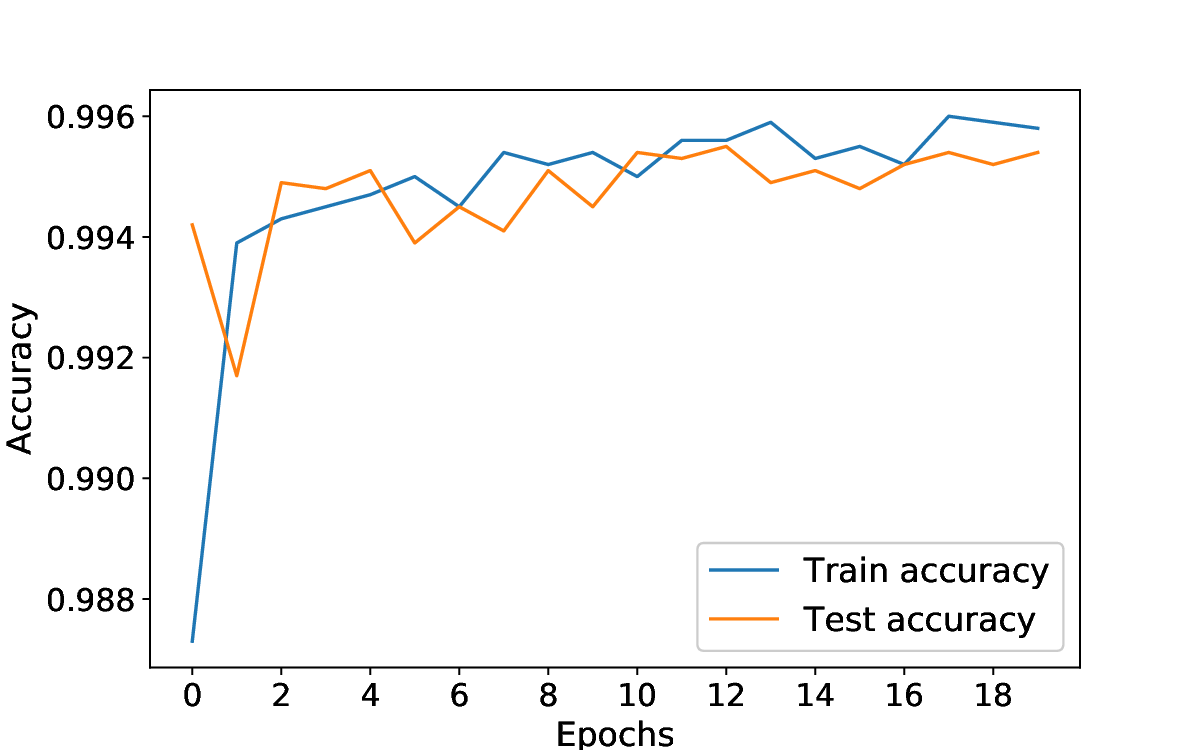}
\includegraphics[height=5cm,width=8cm]{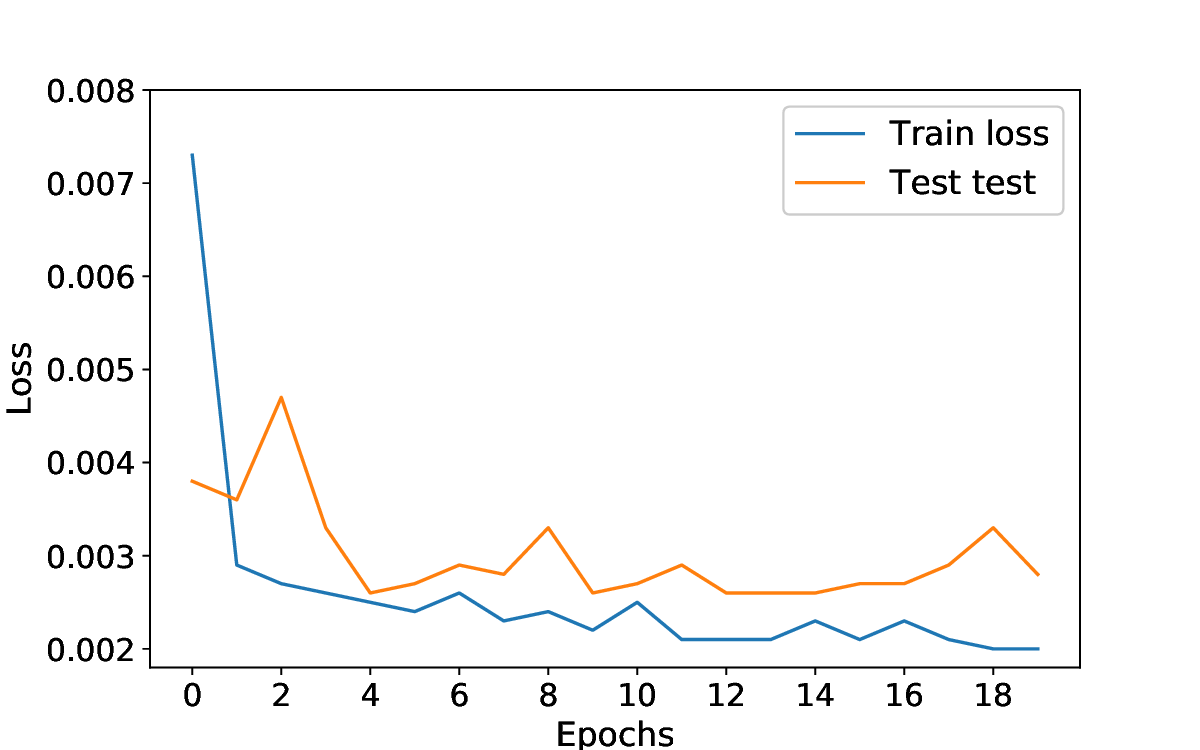}
\caption{\label{fig:acc_loss}
The left is the train/test accuracy and the right is the corresponding loss.}
\end{figure*}

We aim to use ResNet to search for galaxy-LAE lens candidates from the DR12 of  BOSS of SDSS-{\uppercase\expandafter{\romannumeral3}}. At beginning, we need to filter out all the ETG spectra from the database to build our predictive data. The BOSS targets have been classified as several classes, such as ``GALAXY", ``SKY", ``STAR". Therefore, we can choose the ``GALAXY" class and reject the targets with subclass of ``STARFORMING", ``BROADLINE" and ``AGN". We then select the target with redshift of $z > 0.3$. These selections result in 1,241,799 recorders. For these recorders, we only use the spectra from 3800\r{A} to 5000\r{A} to build our predictive spectra. Notice that we do not use the same wavelength range as BELLS GALLERY (3600\r{A} $<\lambda<$ 4800\r{A}), because the spectra in the wavelength from 3600\r{A} to 3800\r{A} is very noisy, which could decrease the accuracy of our ResNet model. 

Our predictive data includes 177 known lens candidates from \cite{2016ApJ...824...86S}. The remaining 10 candidates found in their work are eliminated because of the redshift limitation. Among the 177 candidates, there are 3 spectra with Ly$\alpha$ lines falling out of the wavelength of 3800\r{A} $<\lambda<$ 5000\r{A}. Therefore, only 174 candidates could be used to test our ResNet model.

Finding lens candidates in a great deal of galaxy spectra is an classification problem. Therefore, we use ResNet to build a classifier which can select out the spectra with emission lines not from the foreground galaxies themselves. Then we use several other subsequent selections similar with previous work to reject the false hits. The quality of the train/test data is critical for the accuracy of the classifier and directly affect the efficiency of the selection. For the train/test data, we require: 1$\rangle$ the train/test spectra are not in the DR12 of the BOSS of SDSS-{\uppercase\expandafter{\romannumeral3}}, but look similar as the spectra in the database; 2$\rangle$ the number of the train/test spectra has to be large enough, usually larger than ten thousands; 3$\rangle$ the train/test spectra have been classified into two groups, one group has no Ly$\alpha$ lines (labeled [0]), and the other has Ly$\alpha$ lines (labeled [1]); 4$\rangle$ for the real spectra in DR12, the spectra with Ly$\alpha$ lines are much less than the spectra without Ly$\alpha$ lines, which makes the data imbalance. We need a similar data imbalance in the train/test data. 

So far, there are no more than 200 galaxy spectra with Ly$\alpha$ emission lines from high redshift, which is far from enough. Therefore, we use the following procedures to build an artificial train/test data set.
\begin{enumerate}
\item {We grid the redshift to small bins ($\delta z=0.0025$), and at each redshift bin we perform a Principal Component Analysis (PCA) to the spectra of ETGs \citep{2012MNRAS.421..314C, 2012AJ....143...90S}. Then we use the first six eigenvectors of the PCA to create some artificial galaxy spectra at the corresponding redshift. This procedure yields totally 150,000 artificial galaxy spectra.}

\item {We add the ``SKY" (from BOSS targets) and noise to the created spectra. Now the spectra look similar as real spectra. We then cut out the spectra ranged from 3800\r{A} to 5000\r{A}.}

\item{In the rest-frame wavelength, we create 10,000 artificial Ly$\alpha$ emission lines using Gaussian profiles. The mean wavelength of the Gaussian profiles is at 1216\r{A} and the Full Width at Half Maximum (FWHM) is randomly obtained from a normal distribution with the mean value of $300km\cdot s^{-1}$ and standard deviation of $200km\cdot s^{-1}$. We then limit the FWHM greater than 0.5\r{A} and the peak of the emission lines greater than $1.5\times 10^{-17}erg\cdot cm^2\cdot Ang^{-1}$. Finally, we randomly shift the emission lines to the wavelength of 3800\r{A} $<\lambda<$ 5000\r{A}.}

\item{We randomly select 10,000 spectra from the 150,000 artificial spectra and add the 10,000 artificial Ly$\alpha$ lines to them, then label them as [1]. The remaining spectra without emission lines are labeled as [0].  The above 1/15 selection 
is to overcome the data imbalance mentioned above.}
\end{enumerate}
%The spectra with Ly$\alpha$ emissions are 1/15 of the total train/test spectra in order to overcome the data imbalance we mentioned above.

Now, we have successfully created 150,000 train/test spectra. The PCA approach we performed in the first step could loss some information of the spectra. We point out that the lost information does not affect our training results. Because, firstly, the first 6 eigenvectors are enough to fit a real spectrum with an acceptable $\chi^2$; secondly, most of the lost information is from noise, after adding the ``SKY" and noise in the second step, the created spectra look almost the same as real spectra. Figure \ref{fig:real_arti} shows two high Signal-Noise Ratio (SNR) spectra with Ly$\alpha$ emission lines. The left is the real spectrum of SDSS J1110+2808 (one of BELLS GALLERY samples), the right is an artificial one with a synthetical Ly$\alpha$ line. They look very similar. Another point need to be mentioned is the Gaussian profile approximations for creating Ly$\alpha$ lines. As we will show in Section 4, the Ly$\alpha$ lines are in fact skewed. However, in this study, we just want to use ResNet to recognise the emission lines with any shapes. The shape judgement will be done by some other subsequent selections. Therefore, using skewed profiles to create Ly$\alpha$ lines is not needed. During the experimental process, we can efficiently find the emission lines with any shapes using Gaussian profile approximations.

\subsection{Our ResNet model}

\begin{table*}
\begin{center}
\caption{\label{tb:para} Selected properties of the 5 galaxy--LAE lens candidate systems.}
\begin{tabular}{l c c c c c c c c}
\hline \hline
Target & Plate--MJD--Fiber & $z_{L}$ & $z_{s}$ & R.A. & Decl. & $m_i$ & Ly$\alpha$ Flux  & Skewness\\
\hline
SDSS\, J111749.50$+$014036.9 & 4731-55656-197 & 0.5840 & 2.1570 & 11:17:49.50 & $+$01:40:36.95 &19.35 & 30.4 &2.17$\pm$0.70\\
\hline
SDSS\, J121650.13$+$500700.2 & 6671-56388-589 & 0.6613 & 2.7894 & 12:16:50.13 & $+$50:07:00.26 &19.83 & 36.7
&5.45$\pm$1.60\\
\hline
SDSS\, J122502.89$-$000907.8 & 3847-55588-309 & 0.4868 & 2.4395 & 12:25:02.89 & $-$00:09:07.85 &19.61 & 12.3
&4.03$\pm$1.83\\
\hline
SDSS\, J142310.55$+$231928.1 & 6013-56074-66  & 0.4717 & 2.4099 & 14:23:10.55 & $+$23:19:28.17 &19.66 & 50.8
&1.72$\pm$0.43\\
\hline
SDSS\, J143428.11$+$470111.5 & 6736-56366-505 & 0.5070 & 2.3913 & 14:34:28.11 & $+$47:01:11.53 &19.25 & 24.2
&2.44$\pm$1.43\\
\hline \hline
\end{tabular}
\end{center} 
\textsc{      Note.} --- Column 1 is the SDSS system name in terms of truncated J2000 R.A and decl. in the format HHMMSS.ss$\pm$ DDMMSS.s. Column 2 provides the plate-MJD-fiber of the spectra. Columns 3 and 4 are the redshifts of the foreground lenses and the background galaxies inferred from the BOSS spectra. Column 5 and 6 is the coordinate in terms of truncated J2000 R.A and decl.  Column 7 is the BOSS-measured \emph{i}-band apparent cmodel magnitude within the $1^{\arcsec}$ fiber. Column 8 is the total apparent flux of the Ly$\alpha$ emission lines in units of $10^{-17}erg\cdot cm^{-2} \cdot s^{-1}$. Column 9 is the skewness of the emission line profiles.
\end{table*}

\begin{figure*}
\centering
\includegraphics[height=6cm,width=4.8cm]{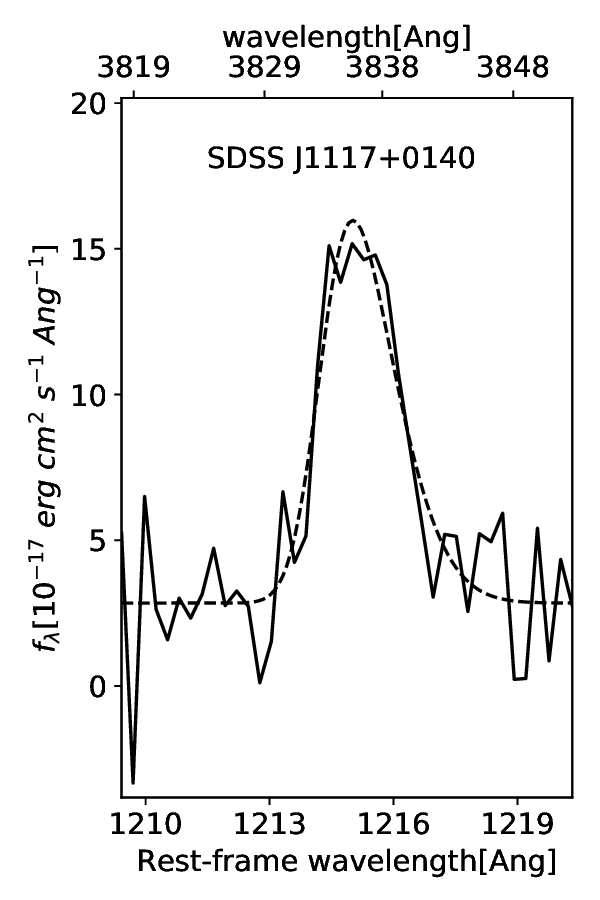}
\includegraphics[height=6cm,width=4.8cm]{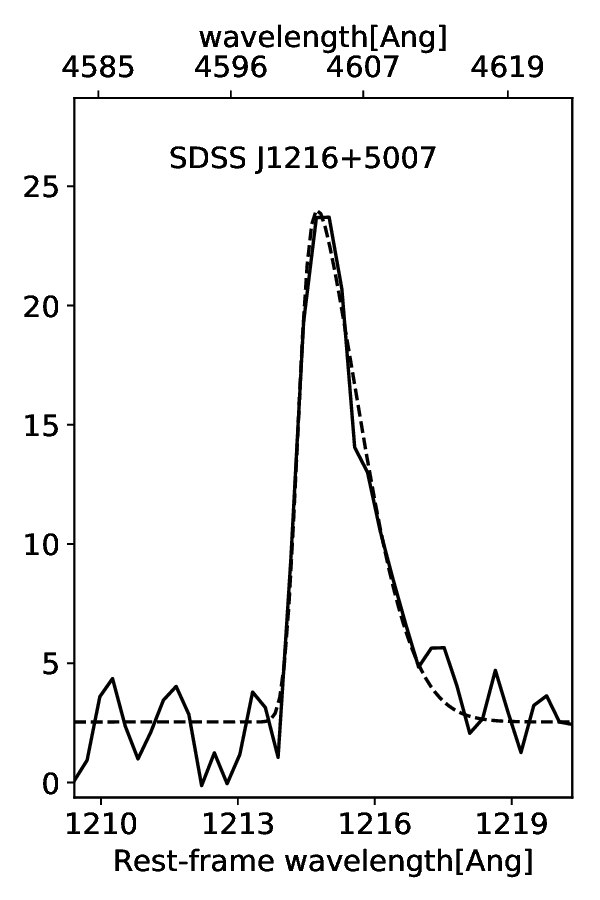}
\includegraphics[height=6cm,width=4.8cm]{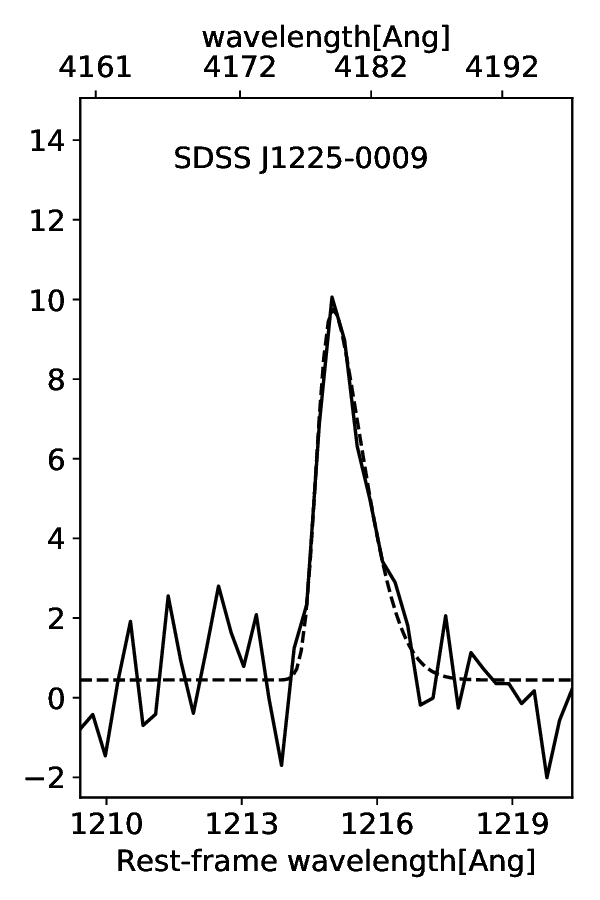}

\includegraphics[height=6cm,width=4.8cm]{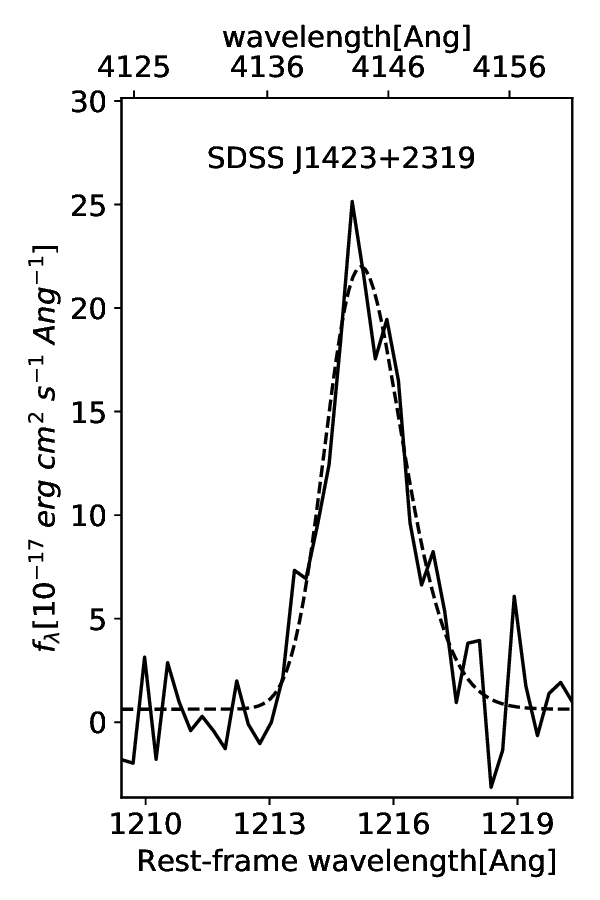} 
\includegraphics[height=6cm,width=4.8cm]{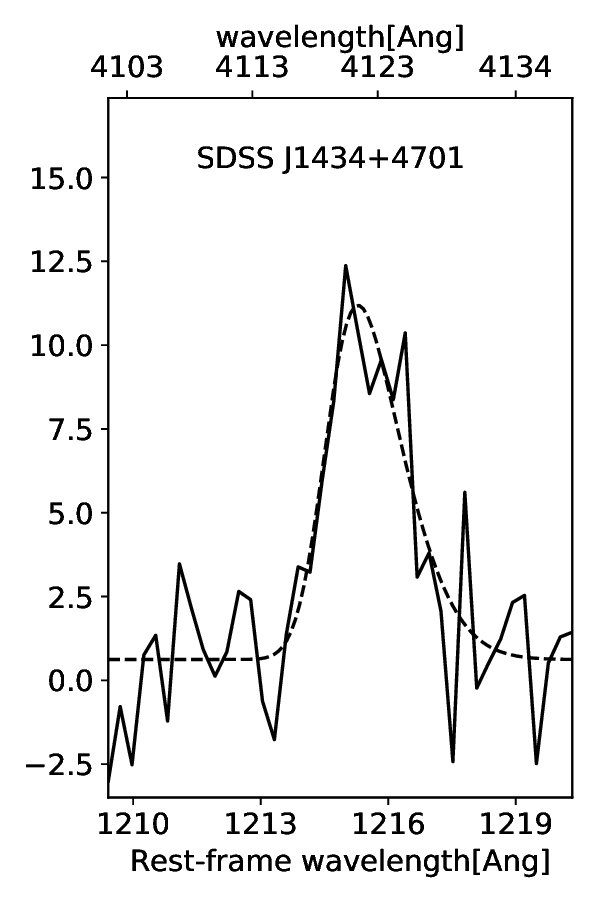}
\caption{Zoomed in views of the Ly$\alpha$ emission lines for the candidates ordered by decreasing apparent Ly$\alpha$ flux, de-redshifted to the rest frame of the LAEs. Dashed lines are the best-fit continuum flux for the Ly$\alpha$ emission lines using skew normal distribution. Note the line profiles with ``blue edge" and ``red tail" are the characteristic of LAEs.}
\label{fig:candidates}
\end{figure*}

Our ResNet model is implemented using the Keras with Tensorflow as its backend. Our model has 8 blocks, 28 convolutional layers, much deeper than plain CNNs. We add a  ``max pooling" layer between any two blocks. A Dropout layer is employed before the fully-connected layer to prevent over fitting. As mentioned before, the samples with label [1] are far less than the samples with label [0], making the data imbalance. Here, we also add another measurement---``Focal Loss"--- to overcome the effect of the data imbalance. The ``Focal Loss" can be written as
\begin{equation}
L_{fl}=\left\{
\begin{aligned}
-(1-\hat{y})^y log \  \hat{y} \ \ \ \   when\  y=1 \\
-\hat y^ylog(1-\hat{y}) \ \ \ \   when\  y=0,\\
\end{aligned}
\right.
\end{equation}
where $y\in \lbrace0,1\rbrace$ is the real label and $\hat{y}$ is the prediction value. The ``Focal Loss" is designed to improve the classification task with extreme data imbalance \citep{2017arXiv170802002L}.

We then use the train/test data (train data/test data = $9:1$) to train our ResNet model. In the training process,  we mask the location of [Ne{\uppercase\expandafter{\romannumeral5}}] 3347, [Ne{\uppercase\expandafter{\romannumeral5}}] 3427, [O{\uppercase\expandafter{\romannumeral2}}] 3727 and [Ne{\uppercase\expandafter{\romannumeral3}}] 3869 lines which may come from the ETGs themselves. The output of our model for each spectrum is the probability to be a lens candidate. We treat the samples with probabilities greater than 0.5 as lens candidates. Figure \ref{fig:acc_loss} shows the accuracy and loss of the train/test data. We obtain a near-perfect test accuracy at 0.9954, the corresponding loss is 0.0028. Also, the completeness is 93.6\%. After successfully training, we apply the model to our predictive data (the emission lines of the ETGs also have been masked). We find all the 21 BELLS GALLERY lens candidates with the possibilities extremely close to 1. For all the 174 known lens candidates in our predictive spectra, we find 161. The discovery rate is 92.5\%. We do not call the discovery rate as completeness because the total number of the lens candidates in our predictive data are unknown.

\section{New strong gravitational lens candidates}
We find 1232 hits using our ResNet model, including 161 candidates found in \cite{2016ApJ...824...86S}. Apart from the existed hits of the 4982 hits in \cite{2016ApJ...824...86S}, we find 536 new hits. Next, we perform several subsequent selections used by \cite{2016ApJ...824...86S} to remove the false hits. First of all, in order to ensure the high quality, we apply a visual inspection to the 536 hits and reject the samples with larger noise or emission lines of low peak flux (the peak is about less than $1 \times 10^{-17}erg\cdot cm^2\cdot Ang^{-1}$ ).  For the left 124 hits, we remove the hits with significant numerical over-densities in both observed wavelength \citep[associate with airglow features, see it in][]{2016ApJ...824...86S} and BOSS target-galaxy rest wavelength (associate with template-subtraction residuals). This step removes 77 spurious hits and left 47. One typical feature of high redshift Ly$\alpha$ lines is the ``blue edge, red tail" profiles (see BELLS GALLERY samples). Therefore, for each spectrum, we use a skew normal profile plus a horizontal line to fit the emission line and the continuum. This fitting is achieved by non-linear least square method.  We use 4 parameters (the Skewness, Mean Value, Standard Deviation and Kurtosis) to describe the skew normal profiles and 1 parameter to describe the horizontal line. All these 5 parameters are varied during the fitting. The errors of the spectra are provided by SDSS. After this selection, 21 samples with skewness parameters bigger than 0 are left. We then remove 16 samples with low-redshift [O{\uppercase\expandafter{\romannumeral2}}] emission lines. In this step, for each sample, we suppose the line is [O{\uppercase\expandafter{\romannumeral2}}] and then calculate the relative positions of [O{\uppercase\expandafter{\romannumeral3}}], H${\alpha}$ and  H${\beta}$ lines. If any of these positions has an emission line, we reject the sample. Finally, we obtain 5 galaxy-LAE lens candidates. The properties of these 5 candidates are shown in Table \ref{tb:para}. We also show the emission profiles and their best fitting skew normal profiles in Figure \ref{fig:candidates}.

\section{Discussion and improvement}
Comparing with \cite{2016ApJ...824...86S}, we find an similar number of galaxy-LAE lens candidates (161+5) in the same database (DR12 of BOSS) using ResNet. The recognition speed of our strategy is much higher than that of theirs. After successfully training the model, we can finish the recognition process within just one and a half hour using the Central Processing Unit (CPU) of a common laptop. By contrast, the method used in \cite{2016ApJ...824...86S} needs about 10 hours to finish a similar recognition. A Graphics Processing Unit (GPU) with lots of Stream Processors can efficiently accelerate the convolutional operations. Although we have not tried the GPU computation, we would expect that the recognition speed could be even faster. This speed advantage would become important in the coming big data age of astronomy. We note that our recognition process does not include the training process. Although we have to spend several hours to train the networks in the beginning, our strategy is still a high efficient one. Once the model has been successfully trained, it can be used to any other predictive data without retraining.

This work is the first attempt to find lens candidates with ResNet based on spectroscopic-selection. Our strategy still has a large improvement space. The first problem is the noise. During the repeating experiments, we find the spectra with too large noise could decrease the accuracy. The second problem is the subsequent selections. Our ResNet can successfully find the spectra with emission lines from different redshifts, but it can not judge whether they are Ly$\alpha$ lines from higher redshift or [O{\uppercase\expandafter{\romannumeral2}}] lines from lower redshift. Therefore, after the ResNet selection, several subsequent selections are needed. If we want to use ResNet to directly find the emission lines with skewness bigger than 0, building the lines with skewed profile and classifying the train data to 3 classes is an approach, the spectra with no emission lines, the spectra with negative skewness lines and the spectra with positive skewness lines. However, this approach still needs subsequent selections and seems unnecessary in this work. An alternative method is using a regression ResNet model to give the redshifts of the source galaxies. This method requires the whole spectrum. We hope the regression model could learn the redshift of the source through the relative position of multiple emission lines. However, to overcome the data imbalance, a  regression model is more difficult than a classification model.

\section{Conclusions}
In this work, we apply the ResNet to search for galaxy-LAE lens candidates based on spectroscopic-selection. We use 140,000 spectra without Ly$\alpha$ lines and 10,000 spectra with Ly$\alpha$ lines to train a 28 layers ResNet model and get a near-perfect test accuracy of 0.9954 with the corresponding loss of 0.0028. After apply our ResNet model to the 1,241,799 predictive spectra from the DR12 of the BOSS of SDSS-\uppercase\expandafter{\romannumeral3}, we find 1232 hits. For all the 174 known candidates in our predictive spectra, we find 161 of them (92.5\%). Finally, we perform several subsequent selections to 536 new hits and find  5 new galaxy-LAE lens candidates. The most obvious advantage of our strategy is the recognition speed. Comparing with previous work \citep{2016ApJ...824...86S}, we find a similar number of lens candidates in the same database with much less time. This work is the fist attempt to find lens candidates using Machine Learning. We will continue to improve this strategy and hope it could become better.

\section{Acknowledgement}
We acknowledge the financial support from the National Natural Science Foundation of China 11573060 and 11661161010. Y.S. has been supported by the National Natural Science Foundation of China (No. 11603032 and 11333008), the 973 program (No. 2015CB857003), and the Royal Society - K.C. Wong International Fellowship (NF170995). GZ acknowledges funding support from the CAS Pioneer Hundred Talent Program Y7CZ181001.

\end{document}